\documentclass[reprint,showpacs,footinbib,citeautoscript,floatfix,nofootinbib]{revtex4-1}

\usepackage{graphicx}
\usepackage{amsmath}
\usepackage{amssymb}
\usepackage{mathrsfs}
\usepackage{bm}
\usepackage{array}
\usepackage{natbib}
\usepackage{soul}
\usepackage{natbib}
\usepackage{xurl}

\newcolumntype {s}[1]{@{\hspace{#1}}} 
\usepackage{color}
\usepackage[usenames,dvipsnames]{xcolor}
\usepackage{wrapfig}
\usepackage[normalem]{ulem}

\begin{document}

\title{Tuneable magnetic behaviour, electronic structure and nitrogen vacancy formation in Gd$_{x}$Sm$_{1-x}$N}

\author{O.~Porat$^{1}$, E.~Joshy$^{1,2}$, J.~D.~Miller$^{1}$  S.~Granville$^{1,2}$ and W.~F.~Holmes-Hewett$^{1,2,*}$,}

\affiliation{$^{1}$Robinson Research Institute, Victoria University of Wellington, P.O. Box 33436, Petone 5046, New Zealand}

\affiliation{$^2$MacDiarmid Institute for Advanced Materials and Nanotechnology, P.O. Box 600, Wellington 6140, New Zealand}

\affiliation{*Corresponding author: W.~F.~Holmes-Hewett, William.Holmes-Hewett@vuw.ac.nz}

\date{October 7, 2024}

\begin{abstract}

\color{black}The rare earth nitrides are the only series of intrinsic ferromagnetic semiconductors where the interplay of spin and unquenched orbital angular momentum provides access to a range of magnetic behaviour. Furthermore, the magnetic properties can be finely tuned through the combination of multiple lanthanide ions in the nitride. \color{black} Here we present a combined computational and experimental study on the electronic and magnetic properties of Gd$_x$Sm$_{1-x}$N and discuss the effect of cation substitution on the internal exchange field and band structure. We find that as the coercive field of Gd$_x$Sm$_{1-x}$N changes over orders of magnitude via cation substitution the internal exchange field changes by $\sim$~20~\%. Control of these material properties is vital in the field of superconducting spintronics. Finally, motivated by an enhanced concentration of nitrogen vacancies in films with higher Sm content, we investigate computationally the formation of nitrogen vacancy defects in Gd$_x$Sm$_{1-x}$N finding that the formation energy is significantly reduced for vacancy sites adjacent to Sm ions rather than Gd ions.

\end{abstract}
\begin{center}
\makebox[\textwidth][c]{
Phys. Rev. Materials $\boldsymbol{8}$, 116201 (2024); DOI: https://doi.org/10.1103/PhysRevMaterials.8.116201~}
\end{center}

\maketitle

\section{Introduction}
\label{introduction}


The ever increasing global energy demand~\cite{Andrae2015} is proving an effective driver in the development of new devices aiming to enhance the efficiency of computation. Demonstrations of fundamentally new types of electronic devices are becoming increasingly more common, among the most promising are spintronic devices which offer lower power and higher density than traditional electronic devices~\cite{Hirohata2020}. Superconducting spintronic devices pair magnetic and superconducting materials and can offer even lower power and faster information transfer~\cite{Mukhanov2011,Tolpygo2007,Linder2015}. These devices use the finite exchange field present in magnetic materials to manipulate the phase of the superconducting wave function, and can be integrated with existing superconducting electronics platforms to enhance functionality. The development of these devices has shown great results, however a lack of suitable materials leads to challenges, particularly regarding superconducting memory systems~\cite{Soloviev2017,Alam2023}. Here we present a study of novel magnetic materials, rare-earth nitrides, where the tuning of the exchange and coercive fields offer advantages pertinent in the field of superconducting spintronics. 

The rare-earth nitrides (\textit{L}nN) series forms in the rock-salt structure, and unlike the other pnictides~\cite{Hulliger1979} its members are mostly ferromagnetic~\cite{Natali2013}. The combination and competition of spin and orbital contributions to the angular momentum and magnetisation can be seen throughout the series to yield rich magnetic behaviours~\cite{Natali2013,Larson2007}. GdN (T$_C$ $\approx$ 69K) ~\cite{Granville2006} is the most studied, and of the magnetic members the simplest, owing to the half filled 4\textit{f} shell. The $^8$S$_{\frac{7}{2}}$ ground state has no orbital contribution to the magnetisation and a total moment of 7~$\mu_B$~per~Gd$^{3+}$ ion. These ions couple strongly to an applied field though the Zeeman interaction $(\mu \cdot B)$ yielding a small coercive field $<100$~Oe~\cite{Ludbrook2009}. SmN with two fewer electrons and configuration $^5$H$_\frac{5}{2}$ has significantly contrasting magnetic properties. In the less than half filled 4\textit{f} shell the spin-orbit interaction dictates that the spin and orbital contributions to the angular momentum are anti-parallel. In the ferromagnetic phase (T$_C$ $\approx$ 27K)~\cite{Meyer2008} the interaction with the crystal field results in a reduction of the expectation value of both $S_z$ and $L_z$ yielding a near compensation of the spin and orbital moments~\cite{McNulty2016}. The net magnetisation is $0.035~\mu_B$~per~Sm~$^{3+}$~ion, reported experimentally~\cite{Meyer2008,Anton2013}. The ferromagnetic state of SmN has further been analysed from a first principles mean field approach~\cite{McNulty2016}. Importantly, there is full ferromagnetic alignment of the 4\textit{f} electrons inter- and intra-ion with a near zero net moment. This results in a very weak Zeeman interaction with an applied field giving a coercive field orders of magnitude larger than GdN, $>~10~T$ at~2~K~\cite{Holmes-Hewett2018}. 

Recent studies of the magnetic properties of Gd$_x$Sm$_{1-x}$N have found that Sm ions substituted into GdN do not interrupt the ferromagnetic order~\cite{Miller2022,Miller2023}. The strong intra-ion 4\textit{f}-4\textit{f} exchange is present and couples the spin on the different cations ferromagnetically. Even as ferromagnetic order is retained, the near zero moment of the Sm$^{3+}$ ions reduces the \textit{net} magnetisation and enhances the coercive field. This tunability has already been utilised in formation of a cryogenic memory element~\cite{Pot2023}.

The electronic properties of the rare-earth nitrides have also been extensively studied. The series in general is insulating with similar band structures when stoichiometric. Most have a valence band maximum at $\Gamma$ formed from N~2\textit{p} states, and an indirect optical band gap at X between the N~2\textit{p} and \textit{L}n~5\textit{d} conduction band minimum, (\textit{L}n for lanthanide)~\cite{Larson2007,Galler2022}. GdN and SmN again are the two most studied members. The seven occupied majority spin 4\textit{f} states of GdN are $\approx$~7~eV below the Fermi energy while the five occupied majority spin 4\textit{f} states of SmN are more widely spread, centred around 6~eV below the Fermi energy. The most variation is found in the respective conduction bands, where GdN lacks any majority spin 4\textit{f} states, while SmN has two unfilled majority spin 4\textit{f} states $\sim~$1~eV above the conduction band minimum, which weakly hybridise with the Sm~5\textit{d}. The intra-ion 4\textit{f}-5\textit{d} exchange lowers the energy of the majority spin 5\textit{d} states in both materials resulting in a spin-split conduction band minimum. 

The rare confluence of insulating and ferromagnetic properties in the ground state has gained the series interest from the superconducting spintronics community. GdN has already been utilised as the weak link in Josephson junctions to form superconducting spin-filter devices~\cite{Senapati2011,Pal2013,Massarotti2015,Caruso2019}, simple memory devices~\cite{Cascales2019}, and recently a superconducting diode~\cite{Sharma2023}. The interest of GdN as a first choice ferromagnetic insulator for application in Josephson junctions has led recently to a thorough analysis of the properties of the electrodynamics of these structures~\cite{Ahmad2020,Ahmad2022}, and even the proposal of a ferromagnetic qubit~\cite{Ahmad2022b}. As for the rest of the \textit{L}nN series only DyN has been explored regarding Josephson junction structures~\cite{Muduli2014}.

To showcase the rich electronic and magnetic behaviour of Gd$_x$Sm$_{1-x}$N in the context of coupled electronic and magnetic behaviours, we present a combined experimental and computational study of the electronic structure, to complement the recent experimental work on the magnetism~\cite{Miller2022,Miller2023}. We discuss band structure calculations and optical spectra of  Gd$_x$Sm$_{1-x}$N in the context of the exchange field, conduction band hybridisation and nitrogen vacancy defect formation. We anticipate that in particular, the control of the coercive and exchange fields we demonstrate will motivate the application of Gd$_x$Sm$_{1-x}$N in the context of superconducting spintronics devices, where control of the exchange field is vitally important for example in switchable 0-$\pi$ Josephson junctions.

In the present manuscript we report orders of magnitude increase in coercive field with decreasing Gd concentration in Gd$_{x}$Sm$_{1-x}$N while the exchange field changes by $\sim$~20$\%$ over the same range. Optical spectroscopy shows a decreasing direct optical bandgap with decreasing Gd concentration which can be understood by considering the hybridisation of the 5\textit{d} electron bands in the cations. Finally we observe an enhanced presence of nitrogen vacancy defects in films with a lower Gd concentration which we understand in the context of a computational study revealing a reduced defect formation energy in Sm rich conditions.

\section{Methods}
\label{Methods}

\subsection{Thin film growth,  optical spectroscopy and magnetometry}
\label{filmgrowth}

Gd$_{x}$Sm$_{1-x}$N thin films were grown in a Riber ultra-high vacuum chamber with base pressure on the order of 10$^{-10}$~mbar~(see reference~\cite{Natali2013} for details). Metallic Sm and Gd were evaporated at a flux on the order of $\sim$~1~$\mathrm{\mathring{A}}$/s in the presence of molecular nitrogen at 2$~\times 10^{-5}$~Torr, the films were grown to $\sim$~100~nm and capped with $\sim$~100~nm of AlN. The ratio of Gd and Sm in the films was controlled by varying the temperature of the cells to increase or decrease their relative flux at the substrate. Films used for optical measurements were grown at ambient temperature to reduce the presence of nitrogen vacancy defects~\cite{Ruck2012}, \color{black} the effect of small temperature variations as a result of passive heating from nearby effusion cells is discussed in section~\ref{energy_section} \color{black}. These films were polycrystalline with a (111) texture, the composition was estimated from the beam equivalent pressures of the deposition sources, which were measured at the substrate location before growth. Films used for magnetic measurements were grown at elevated temperatures on both LAO and AlN templated Si substrates, \color{black}these films are epitaxial and details of the growth of rare-earth nitrides on these substrates are reported in references~\cite{Chan2016,Anton2023,Miller2022}\color{black}. The composition of the films used for magnetic measurements was determined by X-ray fluorescence measurements. 

Optical transmission and reflection measurements were conducted at ambient temperature between energies of 0.01~eV to 4~eV in a Bruker Vertex 80v Fourier transform spectrometer. Reflection measurements were referenced using an Al film, and the results then adjusted for the finite reflectivity of Al~\cite{Ehrenreich1963}. The optical measurements were modelled using the software package RefFIT~\cite{RefFit}, as described in Ref~\cite{Holmes-Hewett2019}, with the resulting optical conductivity presented here. Measurements of the magnetic moment were conducted in a Quantum Design SQUID magnetometer with the field applied in the plane of the film. All magnetic data presented were collected at 5~K. 

\subsection{Density functional theory calculations}
\label{DFT}

Density functional theory (DFT) based calculations were undertaken using Quantum ESPRESSO~\cite{QE,Cococcioni2005} and rare-earth pseudo-potentials developed using the rare-earth nitrides~\cite{Topsakal2014}. Self-consistent calculations on the primitive cell were completed using a $k$-mesh with $10\times10\times10$ divisions, while super-cell calculations were on a $4\times4\times4$ division \textit{k}-mesh. The wave function and charge density cut-off energies were 50 Ry and 200 Ry respectively for all calculations. For all calculations the placement of the various Gd and Sm ions on the rock-salt cation site was randomly determined.

The 4$f$ electrons of the \textit{Ln}N series are strongly correlated and thus require careful treatment beyond the traditional DFT methods~\cite{Larson2006,Larson2007}. In the basic DFT (i.e. LSDA) calculations the 4$f$ states are found at or near the Fermi energy for most of the  stoichiometric \textit{Ln}N. In reality the strongly correlated nature of these electrons pushes the filled states below and unfilled states above the Fermi energy. This physics can be approximated using the DFT+$U$ method where the behaviour of the correlated orbitals is determined by an adjustable parameter $U$. In the present study two $U$ parameters are used, as described first in reference~\cite{Larson2006}, one to account for the strongly correlated 4$f$ states ($U_f$), and a second applied to the 5$d$ states ($U_d$). Selection of the Hubbard parameters is guided by recourse to experimental results, and is discussed in references~\cite{Holmes-Hewett2021} and~\cite{holmes-hewett2023}.

Calculations of the cohesive energy were conducted on 54 atom super-cells of (Gd$_x$Sm$_{1-x}$)N of varying lanthanide composition $x$. The cohesive energy is calculated by taking the difference between the calculated energy of the crystal and the constituent atoms 

$$E_{cho}=n_{Gd}E(Gd)+n_{Sm}E(Sm)+n_{N}E(N)-E(GdSmN),$$ 

\noindent where $n_i$ is the number of each species in a given solid solution, $E(i)$ is the energy of the isolated atoms, and $E(GdSmN)$ is the energy of the given solid solution. The formation energy $E_f$ of nitrogen vacancies was calculated by comparing the energy of the pristine cell with no vacancy defect and the energy of a defective cell containing a vacancy. The difference of the defective cell energy $E_d$ to the energy of the pristine cell $E_p$ plus half the energy of an isolated nitrogen molecule $E(N_2)$,

$$E_f=E_d-E_p+\frac{1}{2}E(N_2)$$

\noindent yields the defect formation energy of a single nitrogen vacancy in nitrogen rich conditions~\cite{Punya2011}, similar to the growth conditions of the thin films in this study. The specific coordination of the nitrogen vacancy (\textit{i.e.}, number and position of Sm or Gd nearest neighbours) was found to be particularly influential on the formation energy. These details are discussed in section~\ref{energy_section}. 

\section{Results}

\subsection{Optical spectroscopy}
\label{optical-section}
\begin{figure}
\centering
\includegraphics[width=\linewidth]{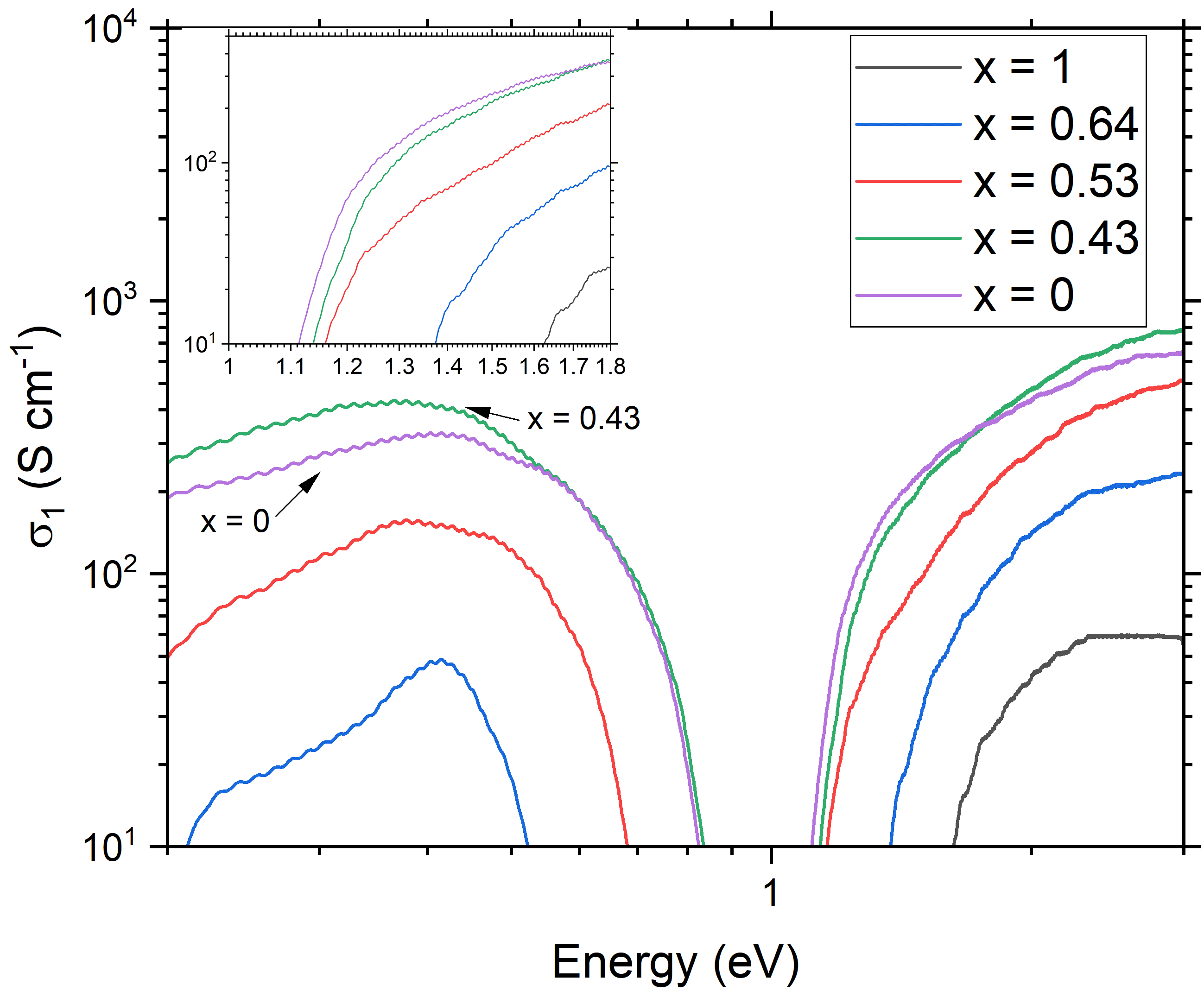}
\caption{Optical conductivity ($\sigma_1$) for all thin films of Gd$_x$Sm$_{1-x}$N based on reflection and transmission measurements, the inset shows the area near the intrinsic band gap.}
\label{optical}
\end{figure}

Reflection and transmission measurements were conducted at room temperature on five Gd$_x$Sm$_{1-x}$N films of nominal composition $x=[0,0.43,0.53,0.64,1]$. The real part of the optical conductivity for the five films is shown in Figure \ref{optical}. Beginning with the GdN film (black), we see that the optical conductivity is essentially zero below $\sim$~1.6~eV, the optical band gap of this film. The gap here is larger than the 1.3~eV which is commonly cited~\cite{Trodahl2007}, although there are reports of GdN with a gap in this range~\cite{Vilela2024}. Importantly this GdN film has no significant contribution to the optical conductivity ($\sigma_1$) below the optical gap (on the scale of Figure~\ref{optical}) indicating a minimal presence of nitrogen vacancy defects~\cite{Holmes-Hewett2020,holmes-hewett2023}. Moving now to the rest of the series, as the Sm content increases ($x$ decreases) the optical edge can be seen moving towards lower energy while the shape remains roughly constant. These observations indicate the 5\textit{d} wave-functions of the Sm and Gd ions hybridise, as these form the final state of this optical transition. In this scenario the optical bandgap will transition from the larger value in GdN to the smaller value of SmN, without the development of additional features above the initial absorption onset. We note that the magnitude of the optical conductivity above the band gap increases with Sm content. A similar enhancement has been observed empirically in both SmN~\cite{holmes-hewett2023} and DyN~\cite{Holmes-Hewett2020} to scale with nitrogen vacancy concentration.

The other clear feature of the optical data is the development of optical transitions and an associated absorption in the mid-infrared region (broad feature in the region below 1~eV) which terminates below the intrinsic optical gap. This feature grows continuously with Sm content, although is slightly larger in the film with $x=0.43$ than the SmN film with $x=0$. This mid-infrared absorption feature has previously been associated with the presence of nitrogen vacancies in rare-earth nitrides~\cite{Holmes-Hewett2020,holmes-hewett2023}. The present data signals that the population of nitrogen vacancy defects is enhanced in films with a greater Sm concentration. We discuss the development of this feature more fully in section~\ref{energy_section} in the context of the calculated defect formation energy. 

\subsection{Electronic Structure}

\begin{figure}[h!!!!!!!!!!!!!!!!!!!!!!!!!!!!!!!!!!!!!!!!!!!!!!!!!!!!!!!!!!!!!!!!!!!!!!]
\centering
\includegraphics[width=0.98\linewidth]{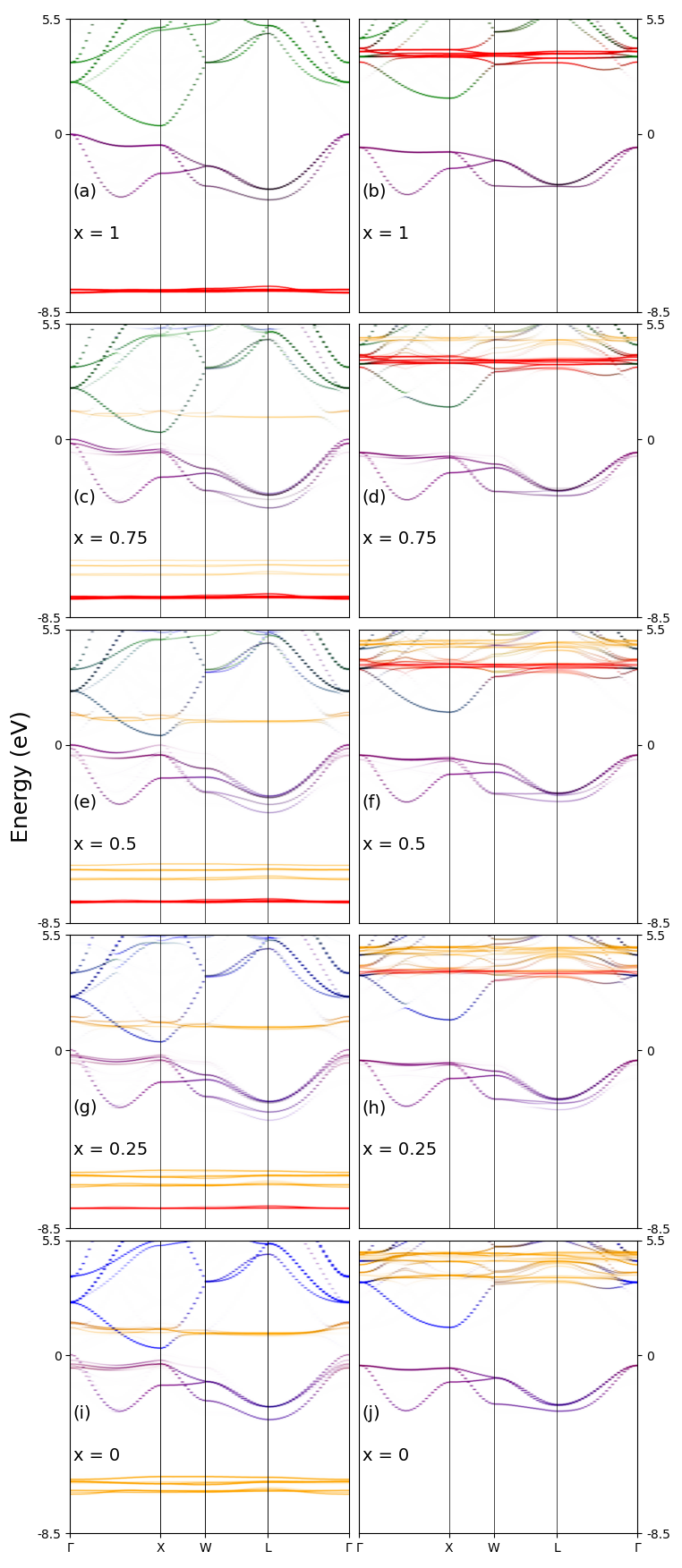}
\caption{Band structure calculations of Gd$_x$Sm$_{1-x}$N for various concentrations of $x$. All left hand panels are majority spin, right hand panels are minority spin. The states are projected onto the atomic orbitals with the following colouring Gd: 4\textit{f}-red, 5\textit{d}-green, Sm: 4\textit{f}-orange, 5\textit{d}-blue, N:~2\textit{p}-purple. The energy range in each panel spans -8.5~eV to +5.5~eV around the Fermi energy.}
\label{array}
\end{figure}

The calculated band structures for various compositions of stoichiometric Gd$_x$Sm$_{1-x}$N are shown in Figure~\ref{array}. Panels~(a) and (b) show the majority and minority bands respectively for $x=1$, or stoichiometric GdN. Here the seven filled Gd~4\textit{f} states (red) are found $\sim$~7~eV below the Fermi energy. The seven unfilled 4\textit{f} states are $\sim$~3~eV above the Fermi energy. The conduction band minimum is majority spin Gd~5\textit{d} (green) at the X point $\sim$~0.4~eV above the Fermi energy. This results in a direct optical gap of $\sim$~0.9~eV, as is found experimentally in the ferromagnetic phase~\cite{Trodahl2007}. The minimum of the minority spin Gd~5\textit{d} band is $\sim$~1.7~eV above the Fermi energy. The difference between the majority and minority spin conduction band minima is twice the exchange splitting ($2E_X$), resulting in $E_X=0.65$~eV for the Gd~5\textit{d} bands, and $E_X=0.2~$eV for the N~2\textit{p} valence band, both at X. 

\begin{figure}
\centering
\includegraphics[width=\linewidth]{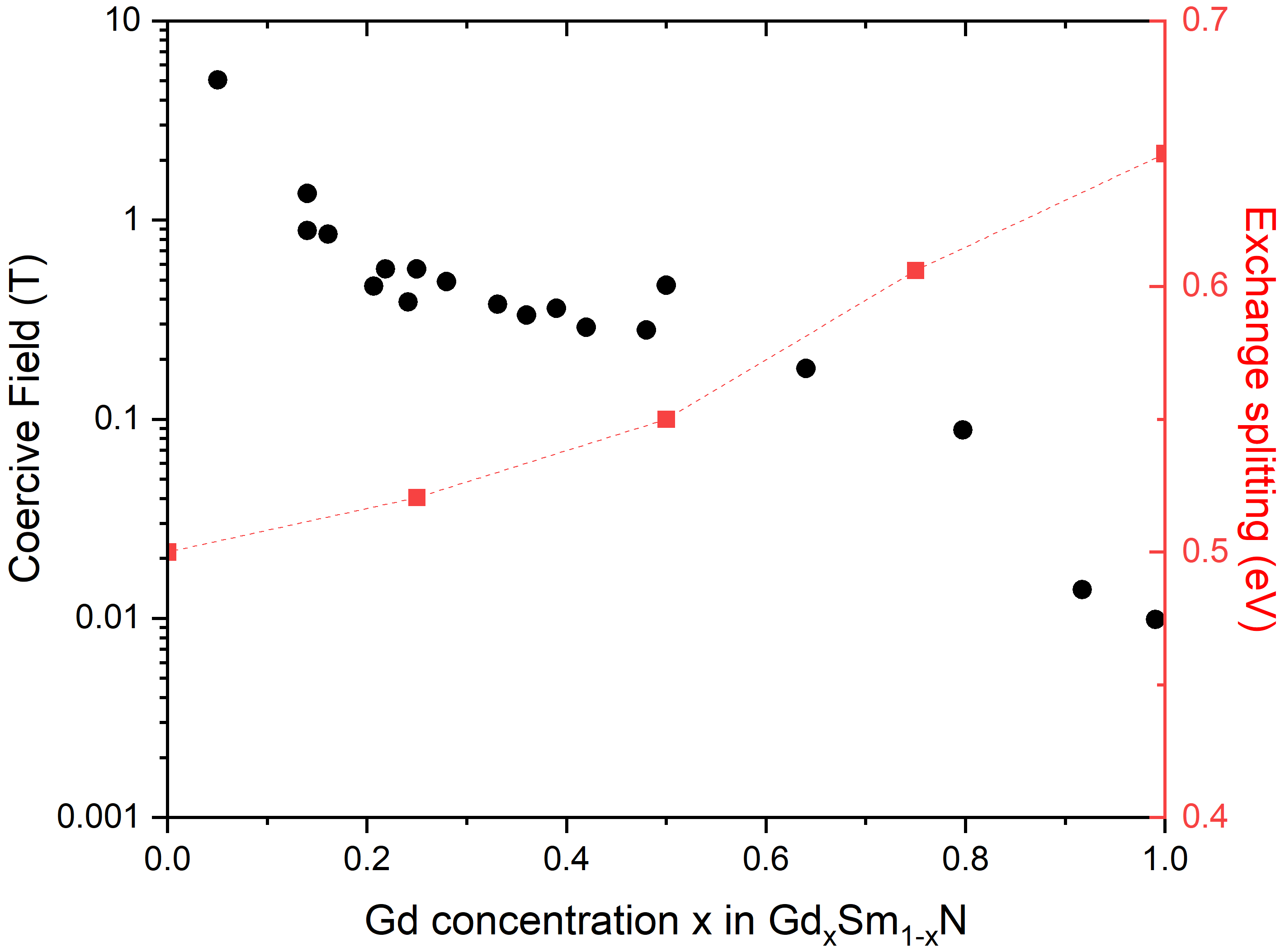}
\caption{Black circles: Experimental coercive field of Gd$_x$Sm$_{1-x}$N thin films of varying Gd content $x$ \color{black} measured at 5~K\color{black}. Red squares: calculated exchange splitting between the majority and minority spin \textit{Ln}~5\textit{d} bands in various Gd$_x$Sm$_{1-x}$N compositions from Figure~\ref{array}.}
\label{Fields}
\end{figure}

Moving now to panels~(i) and (j), here the band structure of stoichiometric SmN is shown. The occupied Sm~4\textit{f} bands (orange) are spread more widely than the 4\textit{f} bands of GdN, centred roughly 6~eV below the Fermi energy. The conduction band minimum is formed from Sm~5\textit{d} states (blue) 0.3~eV above the Fermi energy, with a 0.6~eV direct optical gap at X. There are now unfilled majority spin Sm~4\textit{f} bands which hybridise with the Sm~5\textit{d} bands $\sim$~1~eV above the conduction band minimum. The minority spin Sm~5\textit{d} bands are 1.3~eV above the Fermi energy, resulting in an exchange splitting in the 5\textit{d} conduction band of~$E_X = 0.5$~eV, a reduction of 0.15~eV compared to GdN. The exchange splitting in the conduction band of the \textit{L}nN is a result of the spin polarisation of the \textit{L}n~4\textit{f} states, and their hybridisation with the \textit{L}n~5\textit{d} states. Under the assumption that this hybridisation is similar in GdN and SmN a first approximation to the ratio of the exchange splitting is the ratio of the spin quantum numbers $S_{SmN}$/$S_{GdN}=5/7=0.71$. This value is close to the ratio of the exchange splitting from the calculations of $0.77$.  

As the band structure transitions from GdN to SmN with $x=[0.75,0.5,0.25]$ the respective 4\textit{f} states can be seen emerging at the energies of the end members GdN and SmN. As both the Sm-Gd~4\textit{f} states are localised in space and separated in energy there is little hybridisation between them. In the Gd rich concentration (Fig.~\ref{array}(c)) the majority spin Sm~4\textit{f} states in the conduction band appear even more localised, with near zero dispersion and minimal hybridisation with the lanthanide 5\textit{d} states. In contrast the Gd-Sm~5\textit{d} states are strongly hybridised. The colour of the states can be seen to transition from green to blue though the series of calculations. This smooth transition between the band structures is represented in the continuous change in exchange splitting which is plotted, along with the experimental coercive field, in Figure~\ref{Fields}. The strong hybridisation is consistent with the reduction of optical band gap discussed above.

\subsection{Magnetic behaviour}

As described in section~\ref{introduction}, the strong inter-ion exchange coupling between the 4\textit{f} states of each species results in ferromagnetic alignment of Gd$_x$Sm$_{1-x}$N solid solutions. The response to an applied magnetic field via the Zeeman interaction varies significantly between Gd and Sm ions in the nitrides. As shown in Figure~\ref{Fields}, for high concentrations of Gd the strong Zeeman interaction between the Gd ions and an applied field dominates the behaviour of the material, resulting in a small coercive field. In this situation the strong Gd-Sm exchange coupling causes the spins of the Sm ions to follow the response of the Gd ions to the applied field. With decreasing Gd concentration the coercive field increases significantly. Now the majority Sm ions in the material only interact weakly with an applied field. The Gd-Sm exchange coupling now locks the position of the Gd spins to that of the Sm ions, resulting in a large coercivity. \color{black} Other factors such as the differing anisotropy of GdN and SmN and crystalline quality can also affect the coercivity. The latter has been specifically noted in GdN and DyN~\cite{Anton2023} but can not account for the orders of magnitude change observed here.\color{black}

The evolution of the exchange and coercive fields with varying cation concentration, shown in Figure~\ref{Fields}, is significant, the former changing by only $\sim$~20\% while the latter spans orders of magnitude. The control of coercive field enables the creation of a hard-soft magnetic pair with near equal exchange splitting with potential applications in superconducting spintronics. For example, with $x=0$ the coercive field is $\sim$~100~Oe and the exchange splitting is 0.65~eV, while for $x=0.8$ the coercive field is $\sim$~5000~Oe and the exchange splitting is 0.6~eV. Such a pair could be imagined as layers in a spin-valve Josephson junction where the roughly matched exchange splitting allow selection of the $0$ or $\pi$ state by changing the relative alignment of the layers~\cite{Blanter2004,Gingrich2016}. A further application is the formation of a fringe field free magnetic Josephson junction using a material with $x=0$, SmN. Here the finite exchange splitting allows manipulation of the phase of the superconducting wave-function, while the self-compensated moment significantly reduces the fringe field of the structure compared to traditional materials.

\subsection{Cohesive energy and energetics of nitrogen vacancy formation}
\label{energy_section}
\begin{figure}
\centering
\includegraphics[width=\linewidth]{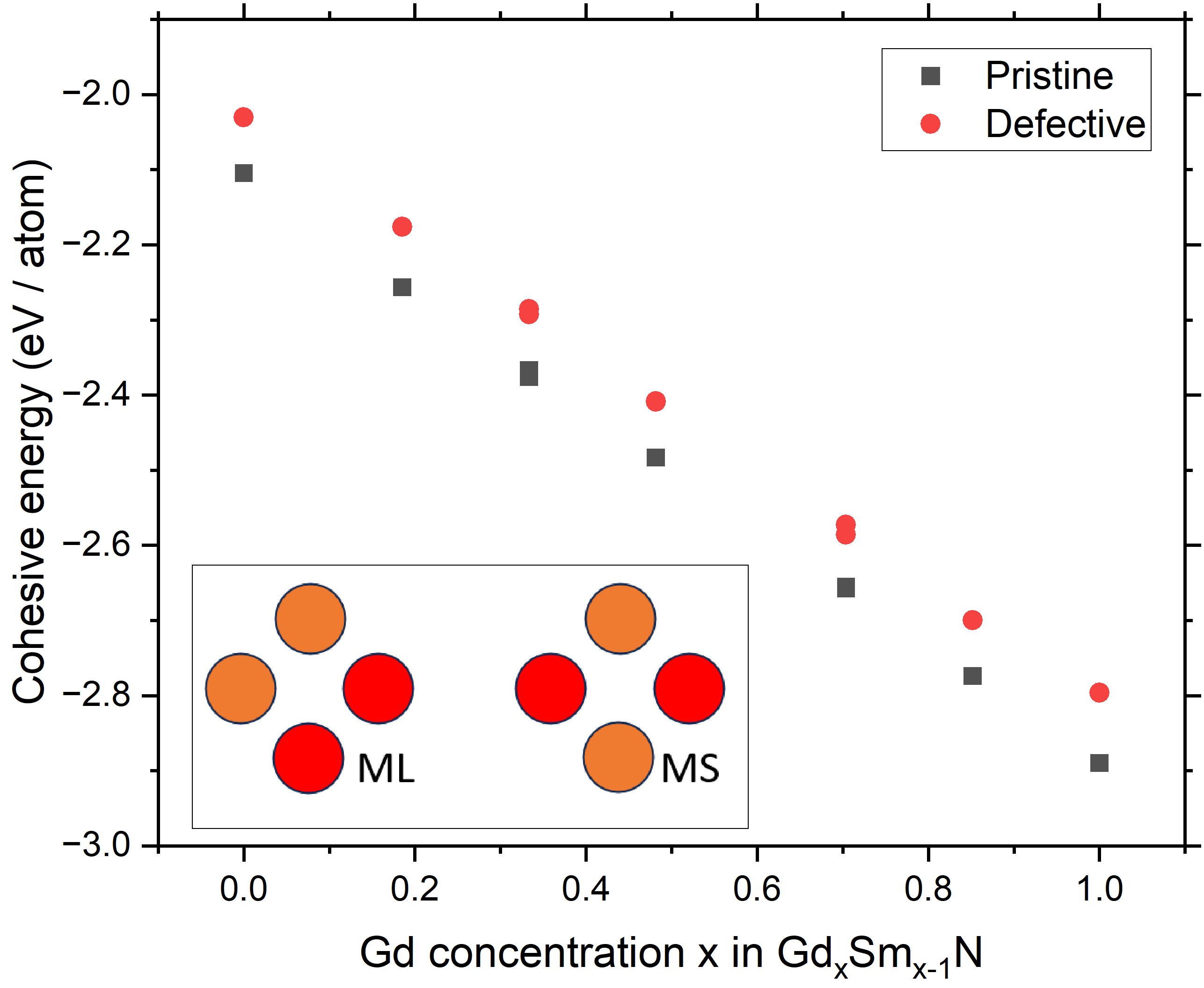}
\caption{Calculated cohesive energy for Gd$_x$Sm$_{1-x}$N as a function of Gd concentration $x$. The schematics show the maximally localised (ML) and maximally separated (MS) configurations of ions around the vacancy, as described in the main text.}
\label{Coh_Energy}
\end{figure}

Motivated by the development of the mid infrared feature in the optical spectroscopy, and the implication of an enhanced population of nitrogen vacancy defects in films with a high concentration of Sm, we have investigated computationally the cohesive energy and defect formation energy of nitrogen vacancies in GdN, SmN and Gd$_x$Sm$_{1-x}$N. The cohesive energy is plotted as a function of concentration in Figure~\ref{Coh_Energy} for both the defective and pristine super-cells, and shows a clear decrease with increasing Gd concentration $x$ indicating that the formation of GdN is energetically favourable over SmN. The $\sim$~20\% difference between these values is quite significant and if not accounted for may impact the compositions of grown films. The cohesive energy is systematically lower for the pristine cells over the defective cells.

In realistic devices comprising \textit{Ln}N films the materials may be doped intentionally or unintentionally with nitrogen vacancies. To help understand this situation in Gd$_x$Sm$_{1-x}$N we have completed a range of calculation for super-cells of various Gd concentration all doped to a nitrogen vacancy concentration of 1/27~$\approx$~3.7~\%, which is representative of heavily doped rare-earth nitride films~\cite{Maity2018,Holmes-Hewett2018,Holmes-Hewett2020}. As mentioned in section~\ref{Methods} the arrangement of the Gd and Sm ions in the solid solution was randomly determined for each calculation. We found the specific coordination of the nitrogen vacancies to be particularly important to the defect formation energy discussed below. In each case the coordination of the nitrogen vacancy in the doped super-cell was chosen to reflect the concentration of the host crystal \textit{i.e.}, a nitrogen vacancy in a (Gd$_{0.7}$Sm$_{0.3}$)N super-cell would have four Gd neighbours and two Sm neighbours. In this case there are furthermore two arrangements of these coordinations, the Sm ions can be maximally separated (MS) on opposite sides of the vacancy, or maximally localised (ML), where the Sm ions are adjacent around the vacancy site. For each calculations with (Gd,Sm) coordination of (4,2), (3,3) and (2,4) there are two configurations, MS or ML, in all cases both were investigated. The inset of Figure~\ref{Coh_Energy} shows the nearest neighbour cations to a nitrogen vacancy in the 001 plane indicating the MS and ML configurations. The cohesive energy was calculated separately for each configuration, but the differences between the respective MS and ML configuration are difficult to see on the scale of Figure~\ref{Coh_Energy}. We now turn to the calculated defect formation energy for nitrogen vacancies, displayed in Table~\ref{Table}.

\label{Doped-calcs}

 \begin{figure*}
\centering
\includegraphics[width=\linewidth]{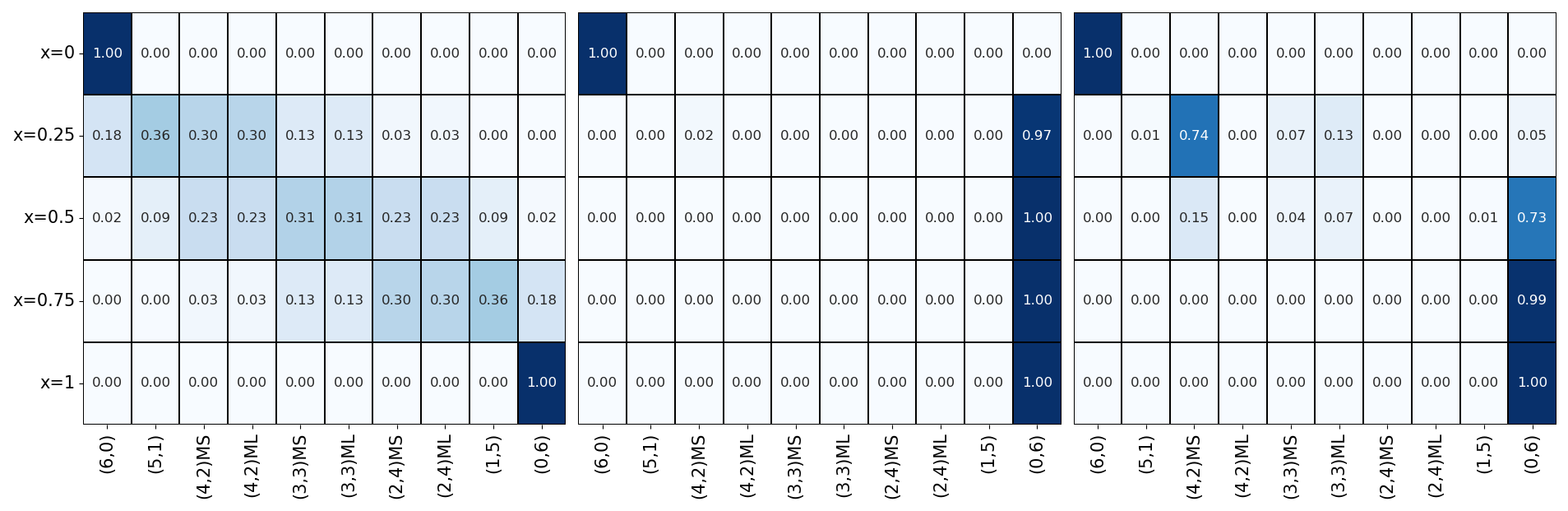}
\caption{Left: Probability of each coordination of a vacancy site based on random chance. Centre: Random chance weighted by the defect formation energy at 300~K. Right: Random chance weighted by the defect formation energy at 1000~K. The three panels show the possible configurations (as in Table~\ref{Table}) along the x-axes and the concentration ($x$) of the Gd$_x$Sm$_{1-x}$N solid solution along the y-axis. The rows are normalised such that the sum along each row is 1.}
\label{Thermal}
\end{figure*}

\begin{table}[]
\begin{tabular}{|l|l|l|}
\hline
\textbf{Gd concentration} & \textbf{Coord. (Gd,Sm)}                 & \textbf{  $E_f$ (eV)  }   \\ \hline
1       &   (6,0)                                         & 2.60                     \\ \hline
0.8     &   (5,1)                                         & 1.48                      \\ \hline
0.7     &   (4,2)MS                                       & 1.09                       \\ \hline
0.7     &   (4,2)ML                                       & 1.90                     \\ \hline
0.48    &   (3,3)MS                                       & 1.22                      \\ \hline
0.48    &   (3,3)ML                                        & 1.16                      \\ \hline
0.33    &   (2,4)MS                                       & 1.46                      \\ \hline
0.33    &   (2,4)ML                                       & 1.54                      \\ \hline
0.18    &   (1,5)                                         & 1.29                       \\ \hline
0       &   (0,6)                                         & 0.71                      \\ \hline
\end{tabular}
\caption{Column 1: $x$ in (Gd$_x$Sm$_{1-x}$)N, Column 2: Coordination of the V$_N$ (Gd,Sm) with ML for maximally localised, MS for maximally separated, Column 3: Formation energy of V$_N$. }
\label{Table}
\end{table}

The general trend for the defect formation energy is that as the Sm coordination of the vacancy site increases the formation energy reduces, this is consistent with the trend of a reducing cohesive energy with increasing Sm concentration seen in Figure~\ref{Coh_Energy}. We interpret this to indicate that vacancies are more likely to occur in Sm rich regions of a given film. There are some points of interest, for example the anomalously low formation energy for the $(4,2)$ MS coordination. One may expect that either the MS or ML configurations would be energetically favourable in general, however, for the (3,3) coordination the ML configuration has a slightly lower nitrogen vacancy defect formation energy.

To extend these results and consider the effect in realistic thin films we take that to first order the defect formation energy is dictated by the nearest neighbours to the vacancy site. In Figure~\ref{Thermal} we have then considered the probability of a given coordination surrounding a vacancy for a film grown with a given concentration $x$ in (Gd$_x$Sm$_{1-x}$)N. The left panel in Figure~\ref{Thermal} shows the non-thermal situation, where the configuration of ions around the vacancy is dictated by random chance alone, the formation energy of the vacancy is not considered. Here as expected the majority weight is along the diagonal. In the central and right panels we have now weighted the random chance of the left panel by the statistical probability of formation $p_i \propto \exp{(\epsilon_i/k_BT)}$, where $\epsilon_i$ is the formation energy of the vacancy for configuration $i$. Now the effect of the formation energy of the different configurations is clear. The significantly lower formation energy for the vacancy site which is coordinated by six Sm ions results in the expectation that for a film grown at room temperature essentially all the nitrogen vacancies will be at sites coordinated by six Sm ions, to within 1\%. The situation is similar, yet not so extreme, at an elevated temperature of 1000~K, approximately the growth temperature of epitaxial GdN and Gd$_x$Sm$_{1-x}$N films~\cite{Ludbrook2009,Miller2022,Miller2023}. Here there is still a clear dominance of the fully Sm coordinated vacancy site, however for more Gd rich concentrations there is at least a few \% of other coordinations present. The effect of the anomalously low formation energy of the (4,2) MS coordination is clear for the Gd rich materials. 

These results can be considered in the context of the optical spectroscopy measurements discussed in section~\ref{optical-section}. Figure~\ref{optical} showed that films with a higher concentration of Sm had a larger absorption feature in the mid infrared, which has been associated with the presence of nitrogen vacancies, implying an enhanced population of nitrogen vacancies in these films. The films used for optical spectroscopy in this study were grown without actively controlling the temperature of the substrate, thus the high temperature of the Gd effusion cell (1400 $^\circ$C) gradually heats the substrate from room temperature to $\sim$~130~$^\circ$C during deposition. The combination of this passive heating and the low defect formation energy for sites coordinated by Sm ions results in a nitrogen vacancy defect population which increases with Sm concentration. 

\begin{figure}
\centering
\includegraphics[width=\linewidth]{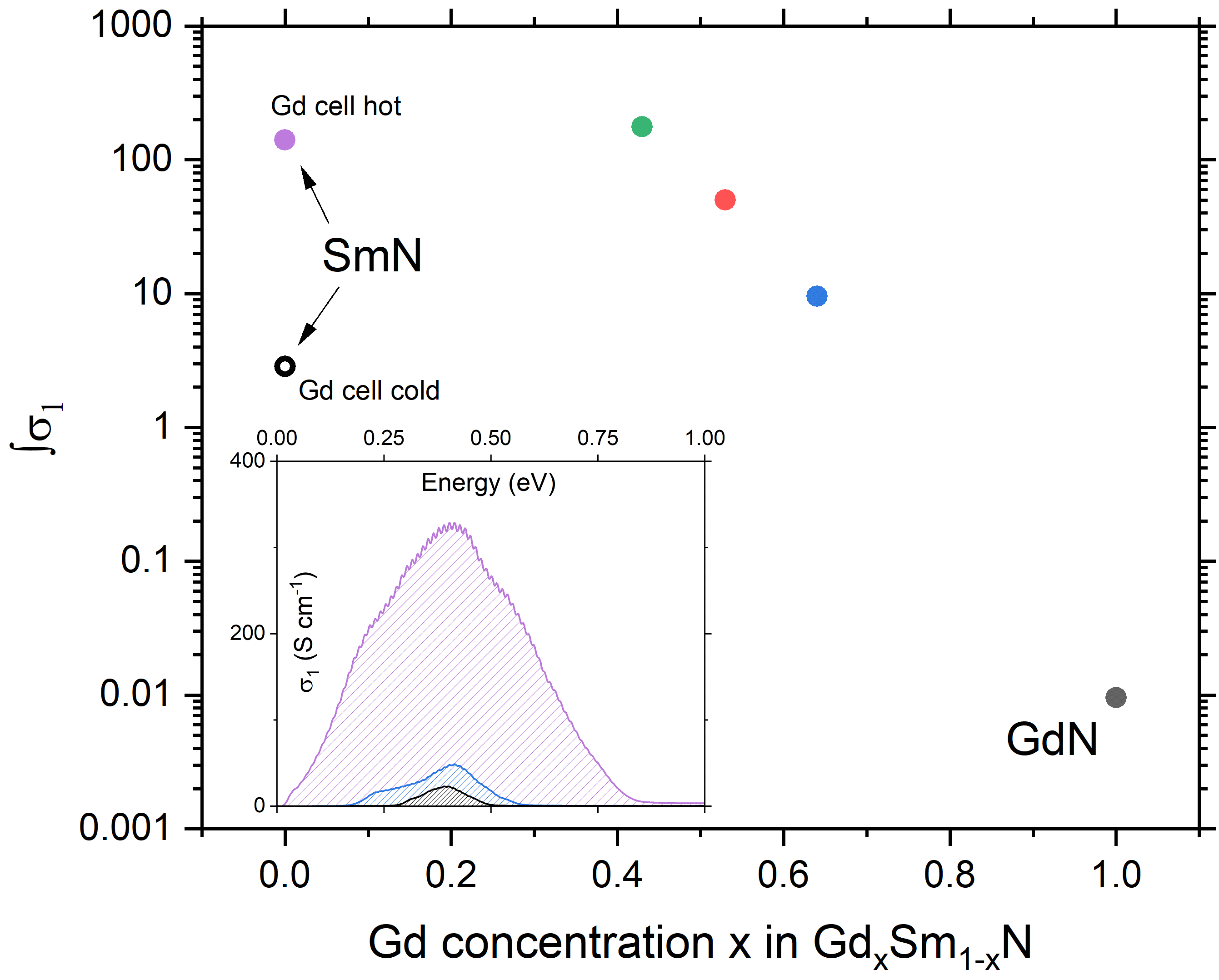}
\caption{Integrated optical conductivity spectra from Figure~\ref{optical} from low energy to 1~eV representing the weight under the mid infrared feature. Two points are present for SmN; these represent films which were grown with (solid circle) and without (open circle) passive heating from the nearby Gd effusion cell. \color{black} The inset shows the area of integration for three films, the SmN films with (purple) and without (black) passive heating, and the film with $x=0.64$ (blue). All solid circle data points are films from Figure~\ref{optical}.\color{black}}
\label{int}
\end{figure}

To observe the sensitivity of nitrogen vacancy formation to temperature we have grown SmN films with and without passive heating from the Gd cell. \color{black} Without passive heating from the Gd effusion cell the temperature of the substrate following thin film growth is lower, only $\sim$~70$^\circ$C\color{black}. We find that the mid infrared absorption feature in the optical spectra, associated with nitrogen vacancy defects, is significantly reduced when the substrate is not passively heated by the shuttered nearby effusion cell. Figure~\ref{int} shows the integrated optical conductivity under the mid infrared absorption feature for all films shown in Figure~\ref{optical}, and in addition a SmN film grown without heating the Gd effusion cell. \color{black} The inset shows the area of integration for three selected films. \color{black}The weight under the mid infra-red absorption feature can be seen to track with Sm concentration for the main series of films, while the SmN film grown without passive heating has a significantly reduced weight implying a reduced population of nitrogen vacancies. 


Finally we comment on the non-monotonic enhancement of the mid-infrared absorption with increasing Sm concentration. As noted in Section~\ref{optical-section} the mid-infrared absorption associated with nitrogen vacancies is larger in the film with $x=0.43$ than the SmN film with $x=0$. As discussed above the latter film was passively heated by the shuttered Gd cell however, this heating was likely reduced compared to the films with non-zero $x$ due to the lack of Gd ions impinging the substrate, resulting in a lower temperature and the formation of comparatively fewer nitrogen vacancies. 

The preferential formation of nitrogen vacancies in Sm rich environments will not only impact the concentration of these defects, but may also influence the nature of electronic transport in electron doped materials. The transport channels in GdN and SmN contrast significantly. GdN has transport behaviours characterised by the Gd~5\textit{d} extended states~\cite{Trodahl2017}, while it is the localised Sm~4\textit{f} states which dominate the electronic behaviours in SmN~\cite{Holmes-Hewett2020}. The preference for Sm coordination of nitrogen vaccines in Gd$_x$Sm$_{1-x}$N may result in electron transport which resembles the localised states in SmN, rather than the extended states in GdN.

\section{Conclusions}

We have reported the optical and magnetic characterisation of Gd$_x$Sm$_{1-x}$N thin films with varying cation concentration, along with density functional theory calculations of the band structure, cohesive energy and defect formation energy of nitrogen vacancies. Our optical spectroscopy studies find that the intrinsic band gap of the materials tracks roughly between GdN and SmN with cation concentration. This observation agrees with the calculated band structure, both indicating that the 5\textit{d} wave functions of the lanthanide ions fully hybridise. We find that the coercive field changes over orders of magnitude with cation composition, while the exchange splitting in the conduction band changes by $\sim$~20~\% over the same range. This tunability could be utilised in multi-layer devices sensitive to the exchange field rather than magnetisation, such as switchable 0-$\pi$ Josephson junctions where matching the integrated exchange field over the subsequent layers is vital. Further the finite exchange field, combined with a near zero moment in SmN may facilitate the manipulation of the phase in a magnetic Josephson junction with a near zero fringe field. Finally our calculations of the cohesive energy and defect formation energy explain the development of a mid infra-red absorption feature which emerges with increasing Sm content in the films. Our data indicates that nitrogen vacancy defects form significantly more readily when fully coordinated by Sm ions, which is more likely to occur in Sm rich films.\\

The data used during this study are available from the corresponding author upon reasonable request.

\section{ACKNOWLEDGMENTS}
This research was supported by Quantum Technologies Aotearoa, a research programme of Te Whai Ao – the Dodd Walls Centre, funded by the New Zealand Ministry of Business Innovation and Employment through International Science Partnerships, contract number UOO2347. We acknowledge additional financial support from the New Zealand Marsden Fund (VUW2106) and the Victoria University of Wellington (FREG). The computations were performed on the R\={a}poi high performance computing facility of Victoria University of Wellington. The authors would like to acknowledge the useful discussions with Ben Ruck, Joe Trodahl and Bob Buckley regarding rare-earth nitrides, and the experimental works, and Martin Markwitz regarding the calculation of defect formation energies.\\ 

Copyright 2024 American Physical Society. This is the accepted manuscript of the following article: Porat, O., Joshy, E., Miller, J.D., Granville, S. and Holmes-Hewett, W.F., “Tuneable magnetic behavior, electronic structure, and nitrogen vacancy formation in GdxSm1-xN”. Physical Review Materials, 8(11), p.116201, (2024).
The final published version is available from [DOI: https://doi.org/10.1103/PhysRevMaterials.8.116201]. This manuscript version is posted with permission for non-commercial scholarly use.

\bibliography{master.bib}

\end{document}